\documentclass[twoside]{article}
\usepackage{graphicx}

\usepackage{lipsum} 

\usepackage[sc]{mathpazo} 
\usepackage[T1]{fontenc} 
\usepackage{microtype} 

\usepackage[hmarginratio=1:1,top=32mm,columnsep=20pt]{geometry} 
\usepackage{multicol} 
\usepackage[hang, small,labelfont=bf,up,textfont=it,up]{caption} 
\usepackage{booktabs} 
\usepackage{float} 
\usepackage{hyperref} 

\usepackage{lettrine} 
\usepackage{paralist} 

\usepackage{abstract} 

\usepackage{titlesec} 
\renewcommand\thesection{\Roman{section}} 
\renewcommand\thesubsection{\Roman{subsection}} 
\titleformat{\section}[block]{\large\scshape\centering}{\thesection.}{1em}{} 
\titleformat{\subsection}[block]{\large}{\thesubsection.}{1em}{} 

\usepackage{fancyhdr} 
\title{\vspace{-15mm}\fontsize{24pt}{10pt}\selectfont\textbf{The rotation of Galaxy Clusters} }

\author{
\large
\textsc{Hrant~M.~Tovmassian}\\[2mm] 
\normalsize 377, W.California, 30, Glendale, CA, 91203, USA \\ 
\normalsize \href{mailto:htovmas@gmail.com}{htovmas@gmail.com} 
\vspace{-5mm}
}

\begin{document}

\maketitle 

\thispagestyle{fancy} 

\begin{abstract}

\noindent 

The method for detection of the galaxy cluster rotation based on the study of distribution of member
galaxies with velocities lower and higher of the cluster mean velocity over the cluster image is
proposed. The search for rotation is made for flat clusters with  $a/b>1.8$ and BMI type clusters 
which are expected to be rotating. For comparison there were studied also round clusters and
clusters of NBMI type, the second by brightness galaxy in which does not differ significantly from the
cluster cD galaxy. Seventeen out of studied 65 clusters are found to be rotating. It was found that the
detection rate is sufficiently high for flat clusters, over 60\%, and clusters of BMI type with dominant
cD galaxy, $\approx 35\%$. The obtained results show that clusters were formed from the huge
primordial gas clouds and preserved the rotation of the primordial clouds, unless they did not have
mergings with other clusters and groups of galaxies, in the result of which the rotation has been
prevented.
\end{abstract}

\begin{multicols}{2} 
\section{Introduction}

The knowledge of the dynamical state of galaxy clusters could provide important constraints
on cosmological scenarios. It is
widely assumed that the dynamics of clusters is mostly governed by infall models and the
theory of caustics (Reg'{o}s \& Geller [1]; van Haarlem et al. [2]; Diaferio \& Geller [3];
Diaferio [4]; Rines et al. [5]. If galactic clusters are formed by hierarchical merging of
groups of galaxies, no rotation of a cluster, as a whole, is possible. However, the formation 
of a cluster from primordial giant gas cloud may not be overlooked. If so, the formed cluster may preserve the rotation of the primordial gas cloud, if afterwards it have not merged with
other clusters and groups. The problem of possible rotation of galaxy clusters has been
discussed by many (Kalinkov [6], Gregory [7], Gregory \& Tift [8], Gregory \&
Thompson [9], Materne \& Hopp [10], Materne, [11] Williams [12] Oegerly \& Hill [13],
Sodre'{e} et al. [14], Biviano et al. [15], Den Hartog \& Katgert [16], Dupke \& Bregman
[17, 18], Tovmassian [19], Burgett et al. [20], Kalinkov et al. [21], Hwang  \& Lee [22]. Though indications of rotation were found in some clusters
[10, 16, 20-22], the general accepted opinion is that galaxy clusters do not rotate. 
The sparse rotating clusters were found among arbitrarily selected cluster samples. Den
Hartog \& Katgert [16] found 13 possibly rotating ones out of studied 72 clusters. Hwang 
\& Lee [22] detected only 13 tentatively rotating clusters among studied 899 Abell clusters. 

For detection the cluster rotation Den Hartog \& Katgert [16] plotted line-of-sight velocity
dispersion against the projected radial distance of the galaxy from the cluster center.
Hwang \& Lee [22] fitted the observed radial velocities $v_p$ of the cluster galaxies with a
function of position angle, $v_p(\theta)$. In both methods it was assumed that the observed
distance of a galaxy from the cluster center is the real distance, and  the galaxy observed at
this position has the velocity corresponding to the rotation model. However, this approach 
could not be applied for all observed galaxies. Tovmassian \& Mntsakanian [23] studied 
3D-distribution of galaxies in clusters and showed that  the number of observed
galaxies over the cluster area with certain radius is sufficiently higher of the number of galaxies
within the sphere of the same radius. For example, the number of galaxies observed over the
small central area of the cluster with radius about five times smaller of the cluster Abell radius,
is by $4\div 5$ times higher of the number of galaxies within the corresponding sphere. Most 
of the observed galaxies here are projected galaxies from the outer spherical shells of the
cluster. Some of them could be located at the cluster border, and their rotational velocities
would be significantly different from the assumed velocities. Also, it was shown [23] that on average about 25\% of galaxies observed over the cluster are
projected galaxies of the cluster environment that have the same velocities, as the cluster
proper members. They will introduce additional errors in the analysis. It follows that  the
possible rotation of the cluster could hardly be revealed by the study of the correlation between
the radial velocity dispersion and the projected galaxy position in the cluster.

I propose a simple method for detecting the cluster rotation, that does not depend on the
projected position of galaxies in relation to the rotation axes. First, in order to minimize the
influence of projected environmental galaxies on the results, it is desirable to study the possibly
small central area of the cluster. I limited the radius of the studied area so in order to have not 
less than about 20 galaxies there. Then, I counted galaxies with velocities lower, $V_l$, and
higher, $V_h$, of the mean velocity $V^*$ of galaxies in the studied area. If a cluster is
experiencing merging, the numbers $n_l$ and $n_h$ of galaxies with velocities lower and
higher of the mean velocity $V^*$ will sufficiently differ from each other. I assume that a 
cluster is in the state of merging, if numbers $n_l$ and $n_h$ differ from each other by more than 1.2 times. If a cluster is in dynamical equilibrium,
the numbers of galaxies with velocities lower and higher of the cluster mean velocity will be
approximately the same (the ratio of their numbers will be smaller than 1.2), at any half of the
cluster image. In a rotating cluster the assumed rotation axes will pass through or be located
close to the adopted cluster center. The numbers of galaxies at two sides of the rotation axes
will be about the same (the ratio of numbers <1.2), but galaxies at one side of the rotation 
axis will have velocities higher of the mean velocity of galaxies in the studied area, and
galaxies at the other side will have velocities lower of the mean velocity. However, due to
interactions between close neighbors, especially in the central dense regions, there could be
member galaxies with not rotational velocities. As a result, the regularities that are
characteristic for a rotating cluster would be somewhat deteriorated. Environmental galaxies
projected over the cluster and interlopers will  mask more the effect of rotation. Anyhow, in a
rotating cluster the majority of galaxies at one side of the rotation axes will move in direction
opposite to the direction of movement of the majority of galaxies at the other side. We assume
that the cluster is rotating, if the portion of galaxies that rotate in the cluster, constitute more
than 60\% of all galaxies at each side of the rotation axes. Hence, by analyzing the distribution 
of galaxies with velocities lower and higher of the cluster mean velocity it will be possible to
detect rotating clusters. 

In this paper I checked the proposed method for detection of the cluster rotation by using four
samples of galaxies. The first sample consists of highly flattened ACO [24] clusters from Strubble and Ftaclas [25] with $f=a/b$ exceeding 1.8. Here $a$
and $b$ are respectively the cluster large and small axis. The high flatness could be evidence
of merging of two clusters. Clusters of high flatness could also be rotating. For comparison I
used the sample of round clusters with $f<1.2$ [25]. The probability of
detection of rotation of such clusters is apparently smaller than in flat ones.

I used also the samples of BMI and non-BMI clusters introduced by Tovmassian \&
Andernach [26], who compared the Abell number count $N_A$ of clusters hosted the cD
galaxy (type I clusters according to Bautz \& Morgan [27]), their
velocity dispersion $\sigma_v$, the peculiar velocity of the cD galaxy and the cluster X-ray
brightness with absolute $K_{s-total}$ magnitude of the cD galaxy, and divided the clusters 
into two types. The clusters, the $K_{s-total}$ 2MASS magnitude of the cD galaxy of which is
by more than $1^m$ brighter than that of the second by brightness galaxy in the cluster, were
classified as BMI. The clusters the $K_{s-total}$ magnitude of the second by brightness galaxy
of which is fainter than the cD galaxy by less than $0.7^m$, were classified as non-BMI (NBMI)
type. Tovmassian \& Andernach [26] suggested that clusters of BMI and NBMI types have
different evolution histories. Clusters of BMI type evolved preferentially without merging with
other clusters. Meanwhile, clusters of NBMI type experienced mergers in history. Therefore,
one could expect that rotating clusters could be found among BMI clusters, whereas hardly NBMI clusters will be rotating.

\section{Analysis and Results}

\subsection{Data}

For our study we used redshifts of cluster members from SDSS-DR9 [28]  that
provides uniform coverage of radial velocities of member galaxies in the whole target area.
Positions of clusters and their redshifts are taken from NED. The galaxies with radial velocities
within $\pm1500$ km s$^{-1}$ from the cluster mean velocity were selected as cluster
members. In order to minimize the influence of projected environmental galaxies, I collected
data for possibly smaller central area of the cluster. Depending on the richness of the cluster
the counts were made within area with radii from 0.25 to 0.75 Abell radius, 
$R_A$\footnote{$R_A $=1.7/$z$ ~arcmin (29)} 
of the cluster. For reliability of the obtaining results, the size of the studied area was chosen so,
to have in it at least 20 galaxies with known redshifts. The compiled lists of the flat and round
clusters contain 18 and 13 clusters respectively. The lists of BMI and NBMI clusters consist of
20 and 16 clusters respectively. The cluster, A1663 is included in two samples: of the round
clusters and of the BMI type clusters. The cluster A2147, is also included in two sample: of the
flat and BMI samples. Hence, the total number of studies clusters is 65.

\subsection{Merging clusters}

In some clusters of all four studied samples the ratio $n_l$/$n_h$ of numbers of galaxies with
velocities lower  and higher  of the mean velocity of galaxies in the studied area exceeds 1.2
or is smaller than 0.8. We assume that these clusters are in the state of merging. The results 
of counts in merging clusters of all four samples are presented in Table 1. In the 1-st column 
of Table 1 the designation of the cluster is presented. In the 2-d column the size of the studied
area as part of the Abell radius $R_A$ is shown. The numbers $n_l$ and $n_h$ are presented
in columns 3 and 4 respectively. In the last column the ratio of the numbers  $n_l$/$n_h$ is
given.

\subsection{Not rotating clusters}		

In some other clusters of all four samples the ratio of the numbers of galaxies moving  towards
the observer and in opposite direction is within $0.8\div 1.20$. These clusters are not 
experiencing merging with other clusters. In all of them it was not possible to determine
a dividing line that could be a rotating axes. Hence, these clusters are not rotating. Their list
is presented in Table 2 analogues to Table 1.

\subsection{Rotating clusters}

In the rest of the studied clusters of all four samples the ratios of the numbers of galaxies
moving towards the observer and in opposite direction are, as in previous group of galaxies,
within $0.8\div 1.2$. For these clusters the possible rotation axis are determined. The number
of galaxies at two sides of the assumed rotation axes in the studied area of
each cluster is about the same, the difference being less than 20\%. The majority of galaxies
at one side of these clusters move in one direction, and the majority of galaxies at the other
side - in opposite direction, that evidences on the cluster rotation. The numbers of galaxies with
rotational movement in these clusters is by $1.5\div3.3$ (with median 2.2) times higher of the
number of other galaxies observed in the cluster area. Therefore, we conclude that these
clusters are rotating. The examples of maps of rotating clusters are presented in Figures 1 and 2 (flat clusters),
Figure 3 (round cluster), and Figures 4 and 5 (clusters of BMI type). The results of counts on rotating clusters are presented in Table 3. In consecutive
columns of Table 3 the following data is presented:  column 1 - the Abell designation of the
cluster; column 2 - the designation of the half of the cluster area for which the information is
presented (W-West, E-East, NE-North-East, etc.); columns 3 and 4 - the number of galaxies
with velocities respectively lower and higher of the mean velocity of galaxies in the studied
area of the cluster. At the upper an lower lanes for each cluster the numbers of galaxies at
corresponding areas are presented.

\section{Conclusions}  

A simple method for detection of rotating clusters is proposed. The essence of the method is
the counts of galaxies with velocities lower and higher of the cluster mean velocity at different
halves of the cluster. The method does not depend on the distance of member galaxies along
the line of sight within the cluster, that affects other methods for search for cluster rotation. The
applied simple method allowed to detect 17 rotating clusters among studied 65, i.e. more than
the quarter of studied clusters are rotating. Note, that the rotation may not be detected for
clusters the rotation axes of which is oriented close to the line of sight. The rate of  rotating
clusters detection is much higher than in other attempts to find rotation (e.g. [16-18; 20-22]). 

The detection rate is incomparably high for flat clusters with $f=a/b>1.8$, which were assumed
to be rotating. In seven out of 18 flat clusters the numbers of galaxies moving in opposite
directions significantly differ from each other. Most probably they are two clusters in the state 
of merging. Out of the rest 11 really flat clusters, seven clusters, i.e. about 64\%, are rotating.
Meanwhile, only two rotating clusters,  $\approx 15$\%, are found among 13 round clusters.
The rate of rotating cluster is also very high among clusters of BMI type, the cD galaxy in which
is brighter than the second by brightness galaxy by more than 1 magnitude. These clusters
preferentially did not have merging in their life, as it was suggested by Tovmassian \&
Andernach [26]. Seven out of the studied 20 BMI clusters, i.e. 35\%, are found to be rotating.
And only one rotating cluster, i.e. $\approx 6$\%, was found among 16 NBMI clusters, which
most probably have experienced mergings in the past [26]. 
The single rotating NBMI cluster is A2147, which is also included in the sample of flat clusters. The high percentage of rotating clusters among clusters of BMI type proves that they
are indeed systems that have not experienced mergings and preserved the rotation of
primordial gas clouds from which they were formed. 

The found high rate of rotating clusters support the opinion that clusters were originally formed
in the rotating primordial gas cloud. Then most of them became reacher in the result of
hierarchical assembly of other groups and clusters of galaxies and, as a result, lost the rotation.

\section{Acknowledgements}

This research has made use of the NASA/IPAC Extragalactic Database (NED) which 
is operated by the Jet Propulsion Laboratory, California Institute of Technology, under 
contract with the National Aeronautics and Space Administration.

\begin{table}[H]	
\caption{Numbers of galaxies with velocities lower, $n_l$, and higher, $n_h$, of the mean velocity $V^*$ of galaxies in the studied area of the cluster, and their ratio in merging clusters.}
\begin{tabular}{lcccc}
\hline
Abell     & $r/R_A$ & $n_l$ & $n_h$ & $n_l/n_h$ \\
\hline
  Flat           \\
\hline
A1187 & 0.50 & 18 & 32 & 0.56 \\ 
A1205 & 0.50 & 27 & 16 & 1.69 \\ 
A1257 & 0.75 & 26 & 16 & 1.62 \\ 
A1371 & 0.50 & 17 & 22 & 0.77 \\ 
A1496 & 0.75 & 14 & 20 & 0.70 \\ 
A2033 & 0.50 & 19 & 25 & 0.76 \\ 
A2175 & 0.60 & 15 & 22 & 0.68 \\ 
\hline
	  Round           \\
\hline
A757 & 0.50   & 14 & 18 & 0.78 \\ 
A1781 & 0.60 & 12 & 18 & 0.66 \\ 
A1890 & 0.40 & 18 & 24 & 0.75 \\ 
A2152 & 0.30 & 12 & 22 & 0.54 \\ 
A2244 & 0.50 & 20 & 16 & 1.25 \\ 
\hline
	 BMI           \\
\hline
A655 & 0.75    & 14 & 19 & 0.74 \\
A1516 & 0.75  & 21 & 29 & 0.72 \\
A1809 & 0.75  & 15 & 21 & 0.71 \\
A1864 & 0.75  & 23 & 16 & 1.44 \\
A2067 & 0.75  & 15 & 27 & 0.66 \\
A2124 & 0.50  & 19 & 27 & 0.70 \\ 
\hline
	  NBMI           \\
\hline
A119 & 0.75   & 33 & 26 & 1.27 \\ 
A1691 & 0.50 & 19 & 26 & 0.73 \\ 
A1991 & 0.40 & 15 & 28 & 0.62 \\ 
A2079 & 0.50 & 15 & 21 & 0.71 \\ 
\hline
\end{tabular}
\end{table}

\begin{table}[H]
\caption{Numbers of galaxies with velocities lower, $n_l$, and higher, $n_h$, of the mean velocity $V^*$ of galaxies in the studied area of the cluster, and their ratio in not rotating clusters.}
\begin{tabular}{lcccc}
\hline
Abell     & $r/R_A$ & $n_l$ & $n_h$ & $n_l/n_h$ \\
\hline
	  Flat           \\
A295 & 0.40   & 28 & 24 & 1.17 \\
A1235 & 0.50 & 14 & 17 & 0.82 \\
A1346 & 0.75 & 20 & 24 & 0.83 \\
A1541 & 0.50 & 24 & 21 & 1.14 \\
\hline
	 Round  \\
\hline
A744 & 0.75   & 13 & 16 & 0.81 \\ 
1663 & 0.50 & 21 & 23 & 0.91 \\
A1750 & 0.50 & 16 & 16 & 1.00 \\
A1927 & 0.75 & 12 & 12 & 1.00 \\
A2149 &0.75 & 19 & 18 & 1.06 \\
A2026 & 0.75 & 19 & 18 & 1.06 \\
\hline
	 BMI 	\\
\hline
A208 & 0.75    & 22 & 18 & 1.22 \\
A279 & 0.50    & 33 & 34 & 0.97 \\
A1302 & 0.75  & 16 & 16 & 1.00 \\
A1663 & 0.50  & 21 & 23 & 0.91 \\
A1925 & 0.75 & 15 & 16  & 0.94 \\
A2029 & 0.50  & 27 & 28 & 0.96 \\
A2244 & 0.75  & 29 & 29 & 1.00 \\ 
\hline
	 NBMI 	\\
\hline
A279 & 0.50   & 33 & 34 & 0.97 \\
A754 & 0.75   & 18 & 15 & 0.12 \\ 
1149 & 0.75    & 17 & 18 & 0.94 \\
A1650 & 0.75 & 27 & 25 & 1.08 \\
A1668 & 0.75 & 15 & 16 & 0.94 \\
A1800 & 0.75 & 25 & 26 & 0.96 \\
A2051 & 0.75 & 13 & 16 & 0.81 \\
A2063 & 0.25 & 24 & 23 & 1.04 \\
A2089 & 0.75 & 30 & 25 & 1.20 \\
A2428 & 0.75 & 16 & 19 & 0.84 \\
A2670 & 0.50 & 25 & 30 & 0.83 \\
\hline
\end{tabular}
\end{table} 	

\begin{table}[H]	
\caption{Rotating clusters.}
\begin{tabular}{lcccc}
\hline
Abell      & $r/R_A$ & Area   & $n_l$ & $n_h$  \\
\hline
		Flat 		\\
\hline
A1035 & 0.50 & NW &  10  & 16 \\ 
	        &      & SW &  18 & 10 \\
A1225 & 0.75 & N    &  12 &  4  \\ 
		&      & S    &   4  & 13 \\
A1362 & 0.75 & E    &   4   &  9  \\
		&     & W   &  8   &   4 \\
A2069 & 0.75 & NW & 11  &   6  \\
               &      &  SE  &  6  & 12  \\
A2110 & 0.75 & NE  &  4   &  7 \\  
		  &    & SW &  8   & 4  \\
A2147 & 0.25 & NW  &  9 & 19  \\
		  &    & SE   & 15 & 12  \\
A2175 & 0.6  & NW  & 11 &  7 \\  
		  &   & SE   &   4  & 15 \\ 
A2197 & 0.25 & NE  & 5   & 11  \\    
		 &     & SW & 10 &  6 \\
\hline
	 Round     \\
\hline
A858 & 0.75  & SE  & 10 &  4  \\   
		 &    & NW & 3  & 10 \\
A1238 & 0.50 &  N  & 15 &  5 \\    
		 &     & W  &  6  & 14 \\
\hline
	 BMI	 	\\
\hline
A85 & 0.30     & NW &  14 & 6  \\  
	          &    & SE   &  7 & 13 \\   
A152 & 0.60   & SE &  13 &   8 \\  
		  &     & NW & 7 & 15 \\ 
A690 & 0.75   &   E   & 8 & 11  \\   
		  &     & W   & 13 &  6 \\ 
A1651 & 0.50  &  N   &   9 & 14  \\  
		  &     & S    & 16 &  8 \\  
A1738 & 0.75 & NE &  5 & 12  \\ 
		  &    & SW & 10 & 6 \\   
A1795 & 0.40 &  N    & 15 &  5  \\
		 &     &  S    &  7 & 13 \\
A1890 & 0.50 & NE  &  6 & 16  \\ 
		  &    & SW & 14 & 6 \\    
\hline
	 NBMI 		\\
\hline
A2147 & 0.25 & NW  &  9 & 20  \\
		 &     & SE   & 15 & 11  \\
\hline
\end{tabular}
\end{table}

\end{multicols}

\begin{figure}
\includegraphics[width=16cm]{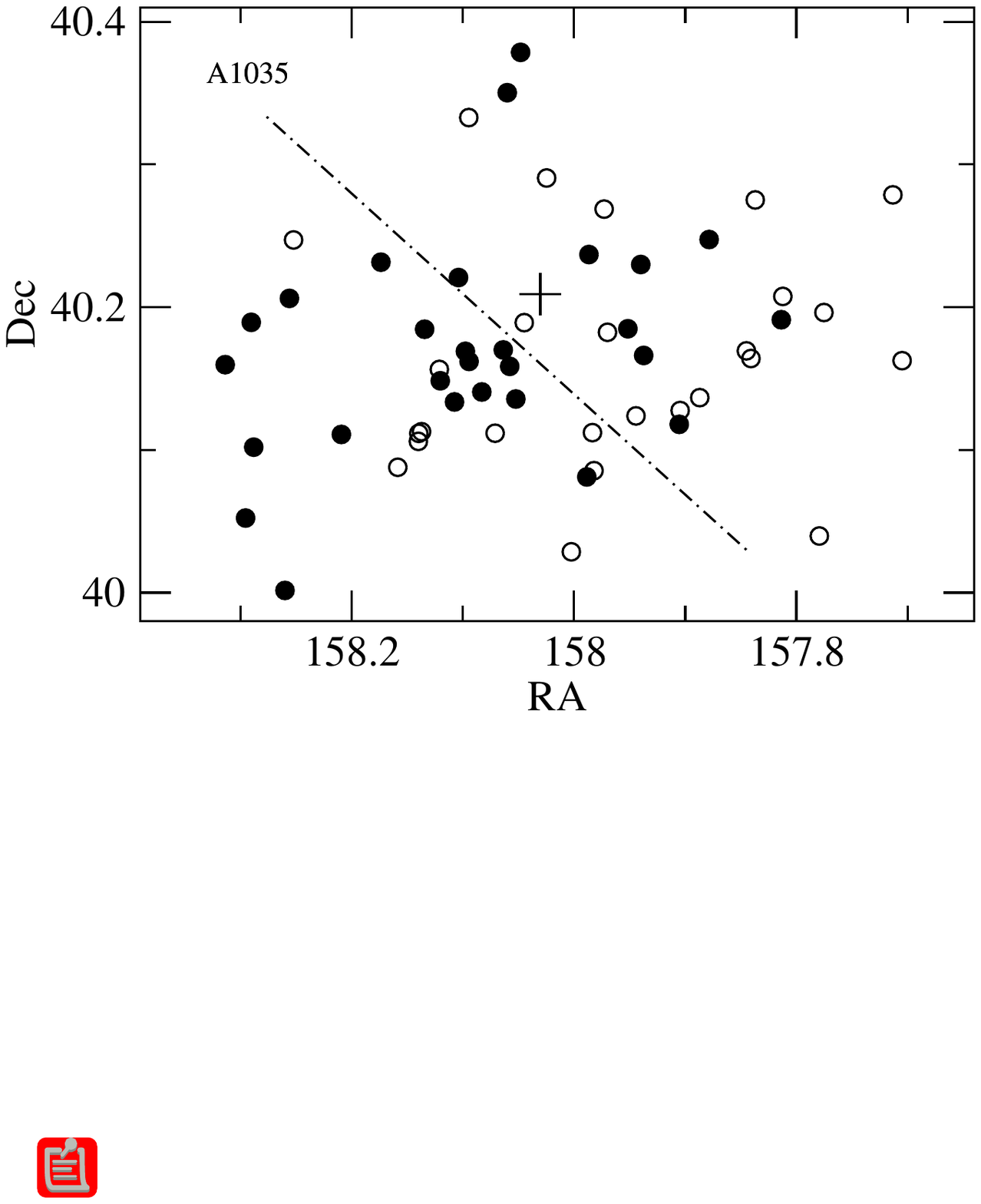}
\caption{The map of the central region ($0.5R_A$) of the rotating cluster A1035. Filled circles 
are galaxies with velocities lower than the mean velocity of galaxies in the studied central 
area of the cluster. Open circles are galaxies with velocities higher than the mean velocity.
Cross is the cluster center according to NED. Dash-dotted line is the assumed rotation axes.
The same marking is used for other maps.}
\end{figure}

\begin{figure}
\includegraphics[width=16cm]{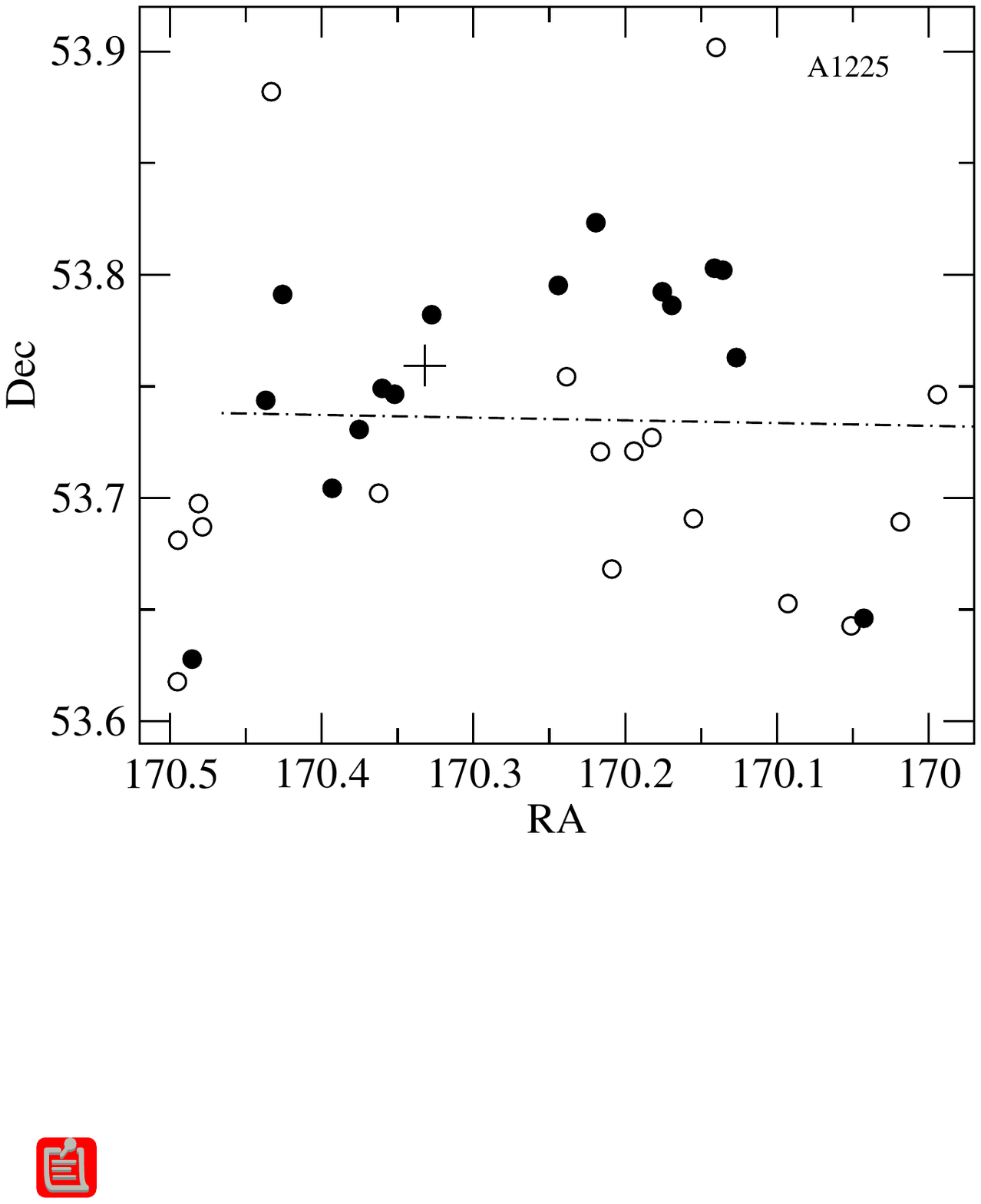}
\caption{The map of the central region ($0.75R_A$) of the rotating cluster A1225.}
\end{figure}

\begin{figure}
\includegraphics[width=16cm]{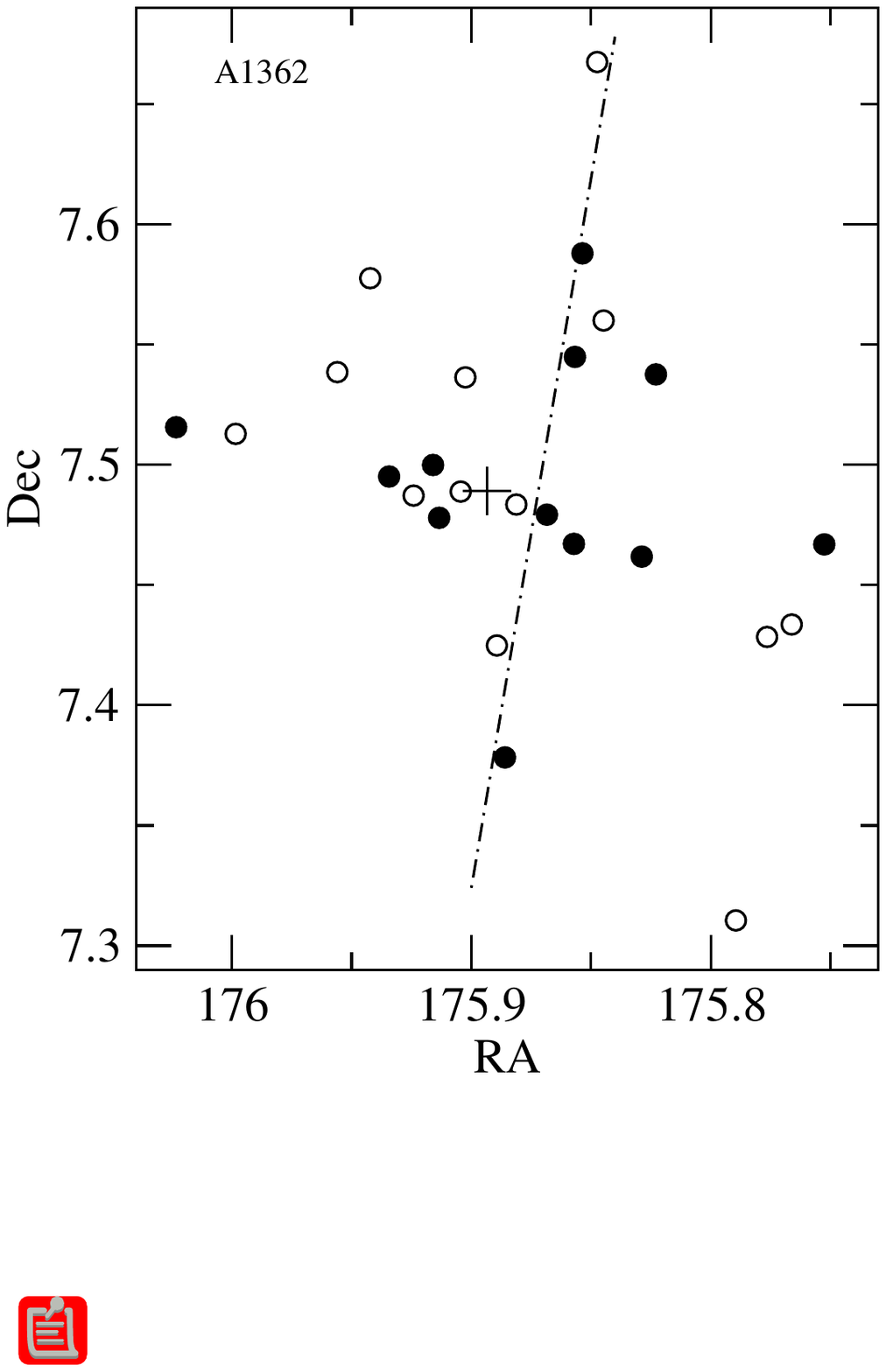}
\caption{The map of the central region ($0.75R_A$) of the rotating cluster A1362.}
\end{figure}

\begin{figure}
\includegraphics[width=16cm]{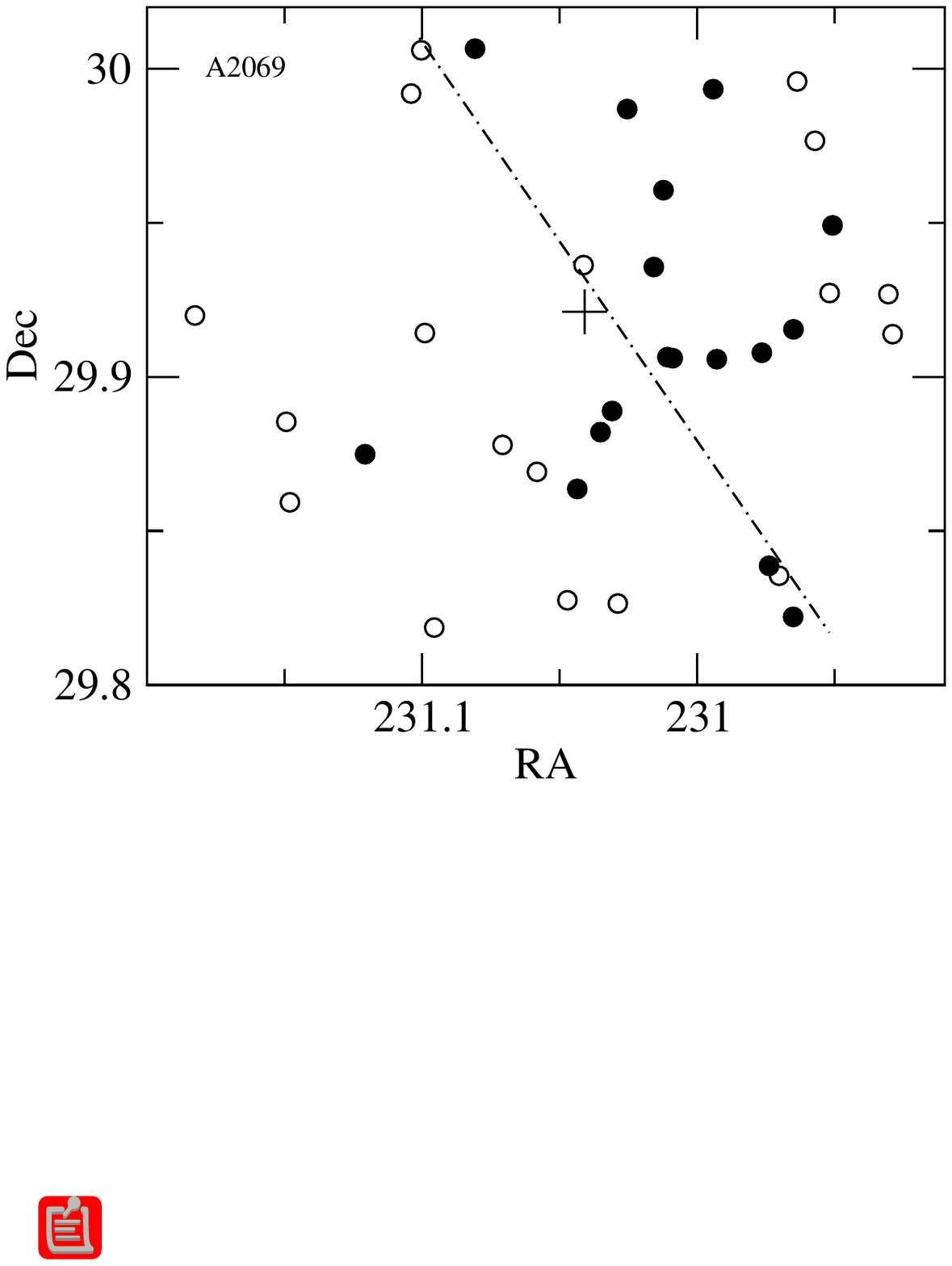}
\caption{The map of the central region ($0.75R_A$) of the rotating cluster A2069.}
\end{figure}

\begin{figure}
\includegraphics[width=16cm]{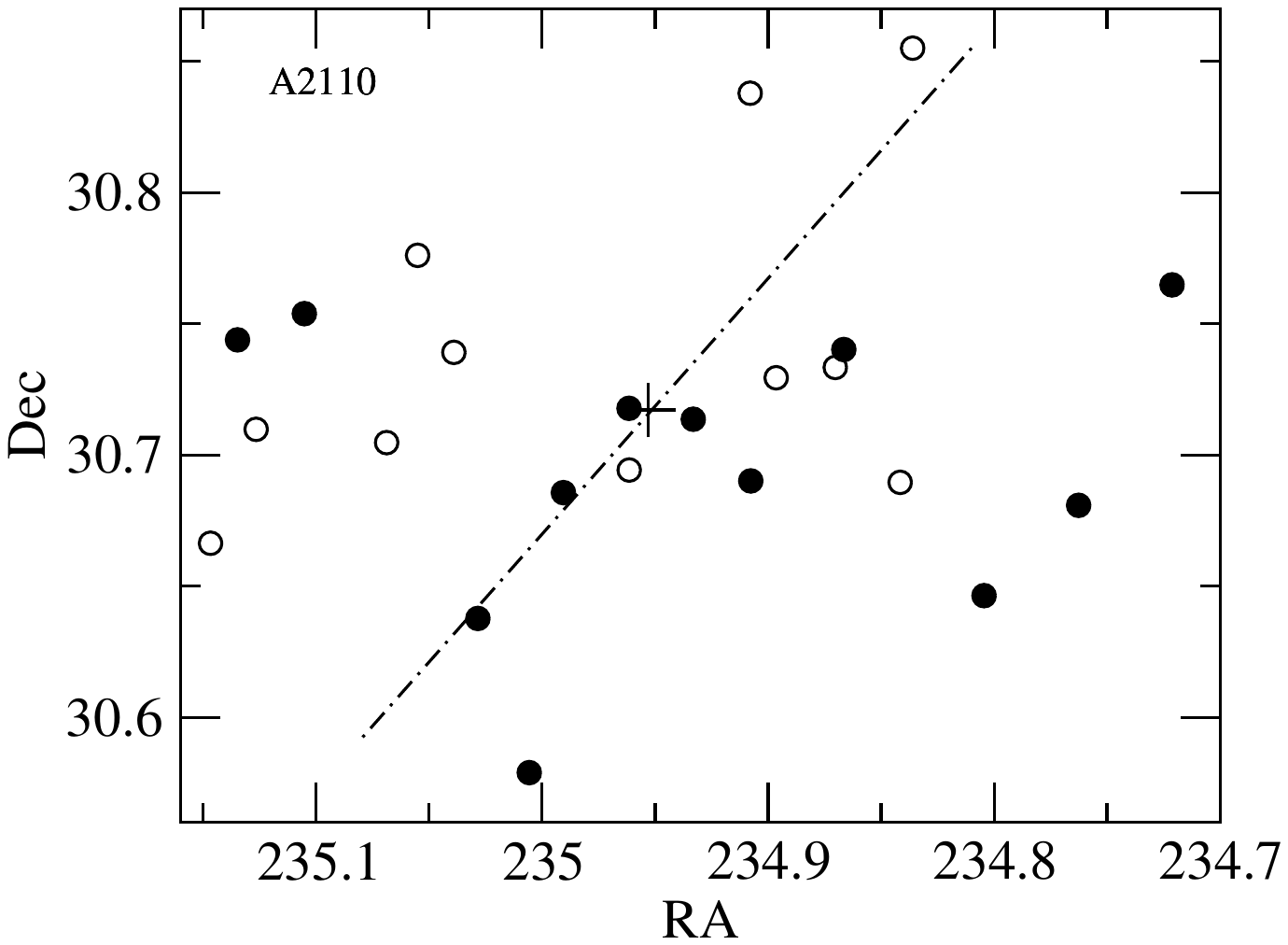}
\caption{The map of the central region ($0.75R_A$) of the rotating cluster A2110.}
\end{figure}

\begin{figure}
\includegraphics[width=16cm]{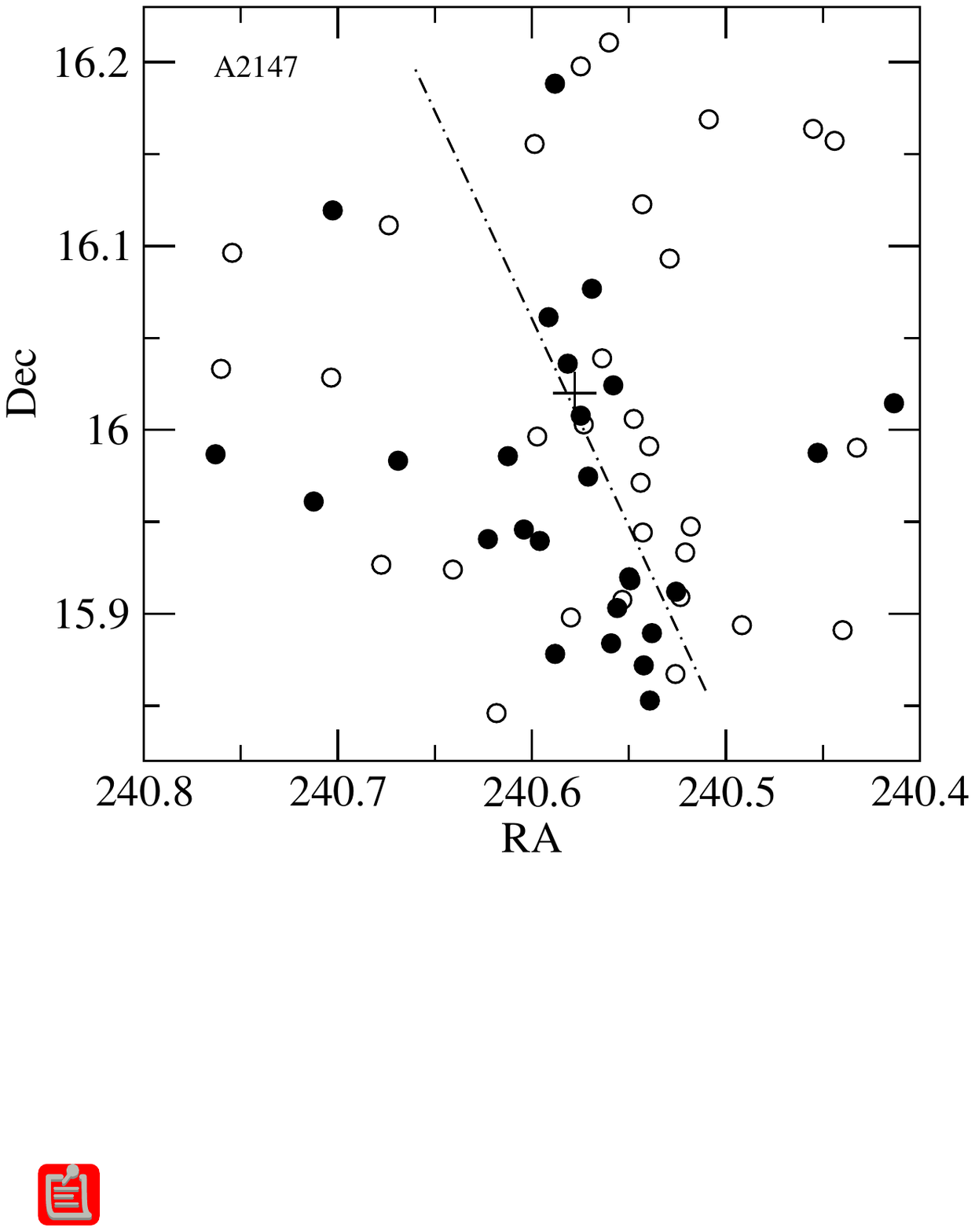}
\caption{The map of the central region ($0.25R_A$) of the rotating cluster A2147.}
\end{figure}

\begin{figure}
\includegraphics[width=16cm]{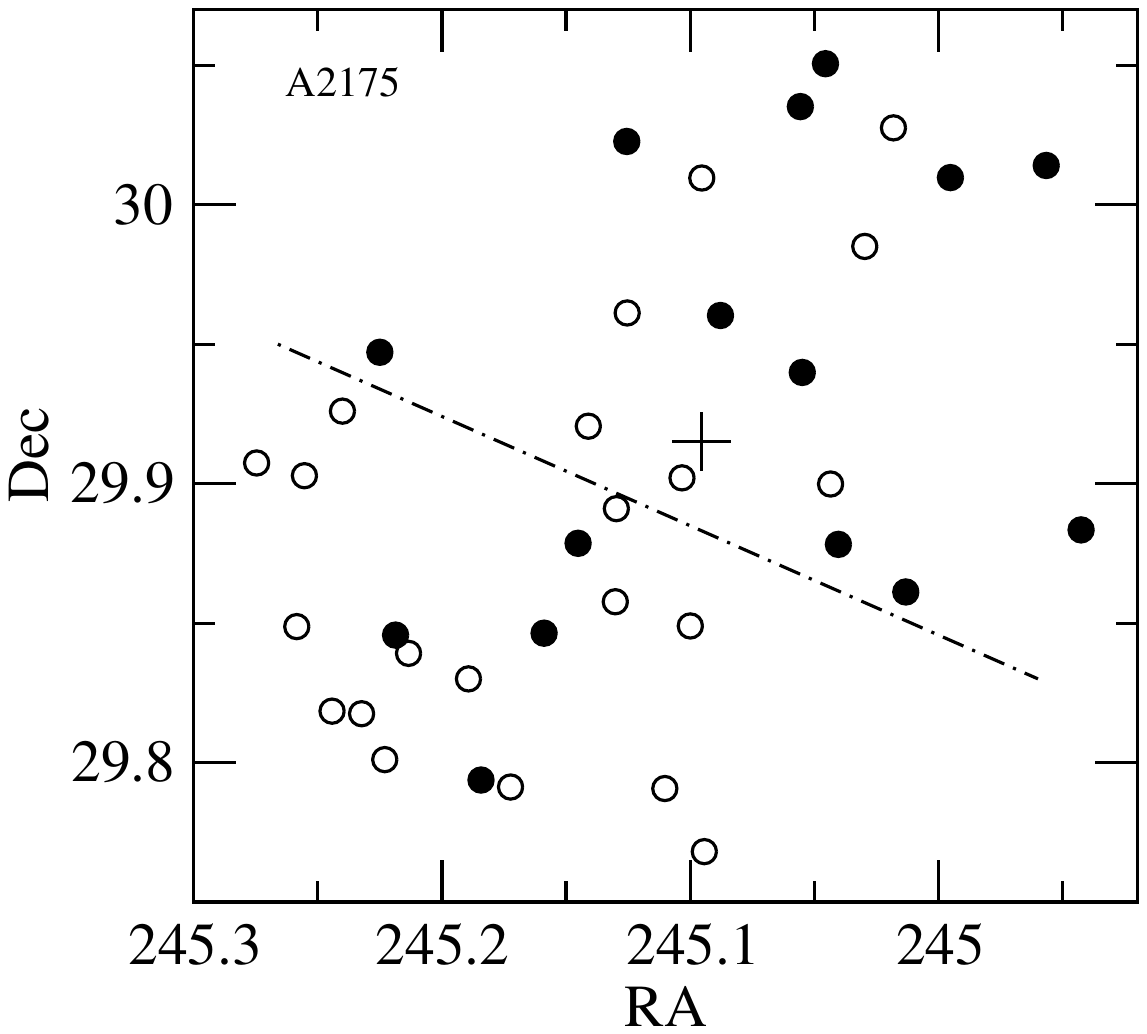}
\caption{The map of the central region ($0.6R_A$) of the rotating cluster A2175.}
\end{figure}

\begin{figure}
\includegraphics[width=16cm]{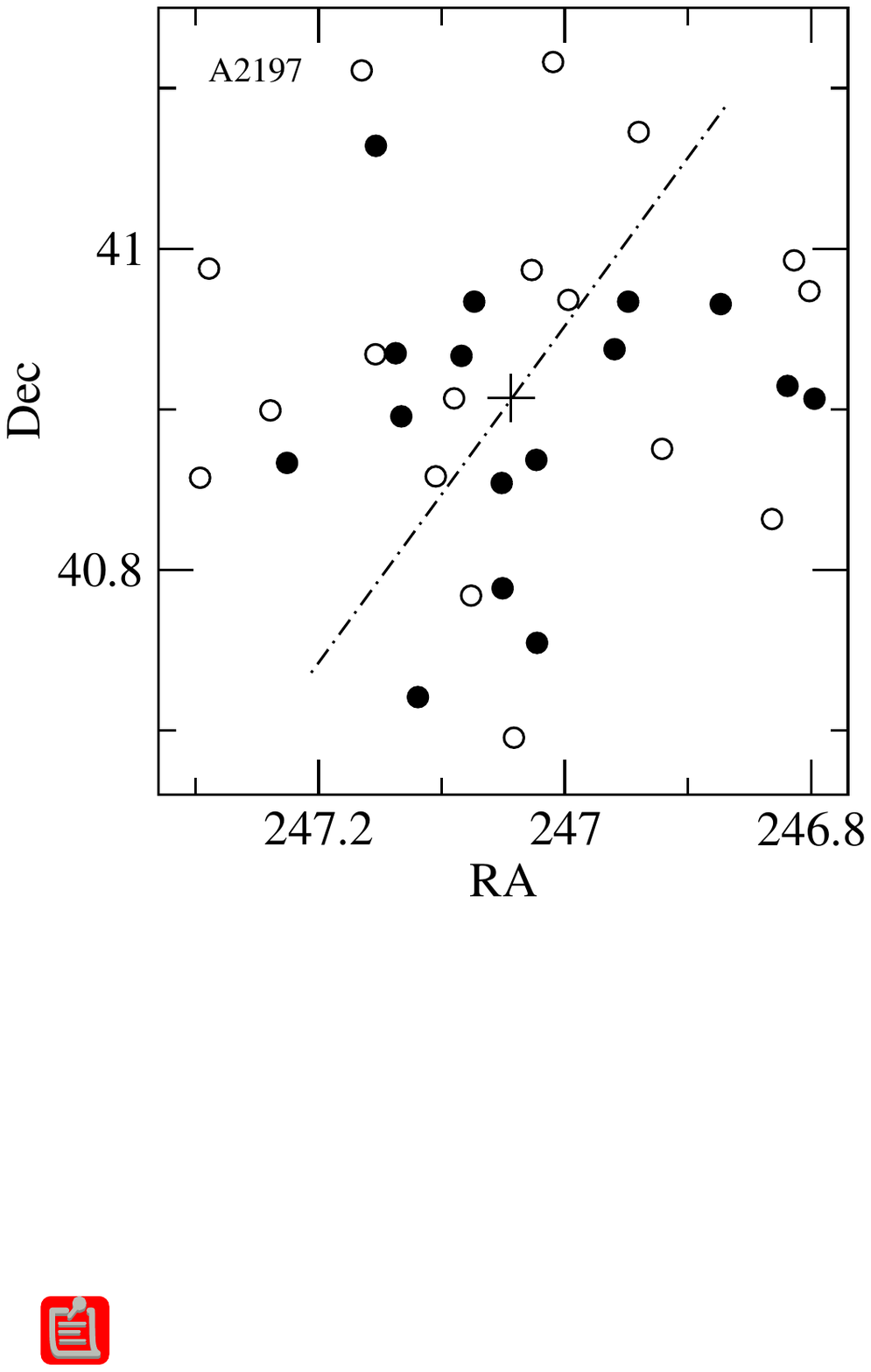}
\caption{The map of the central region ($0.25R_A$) of the rotating cluster A2197.}
\end{figure}

\begin{figure}
\includegraphics[width=16cm]{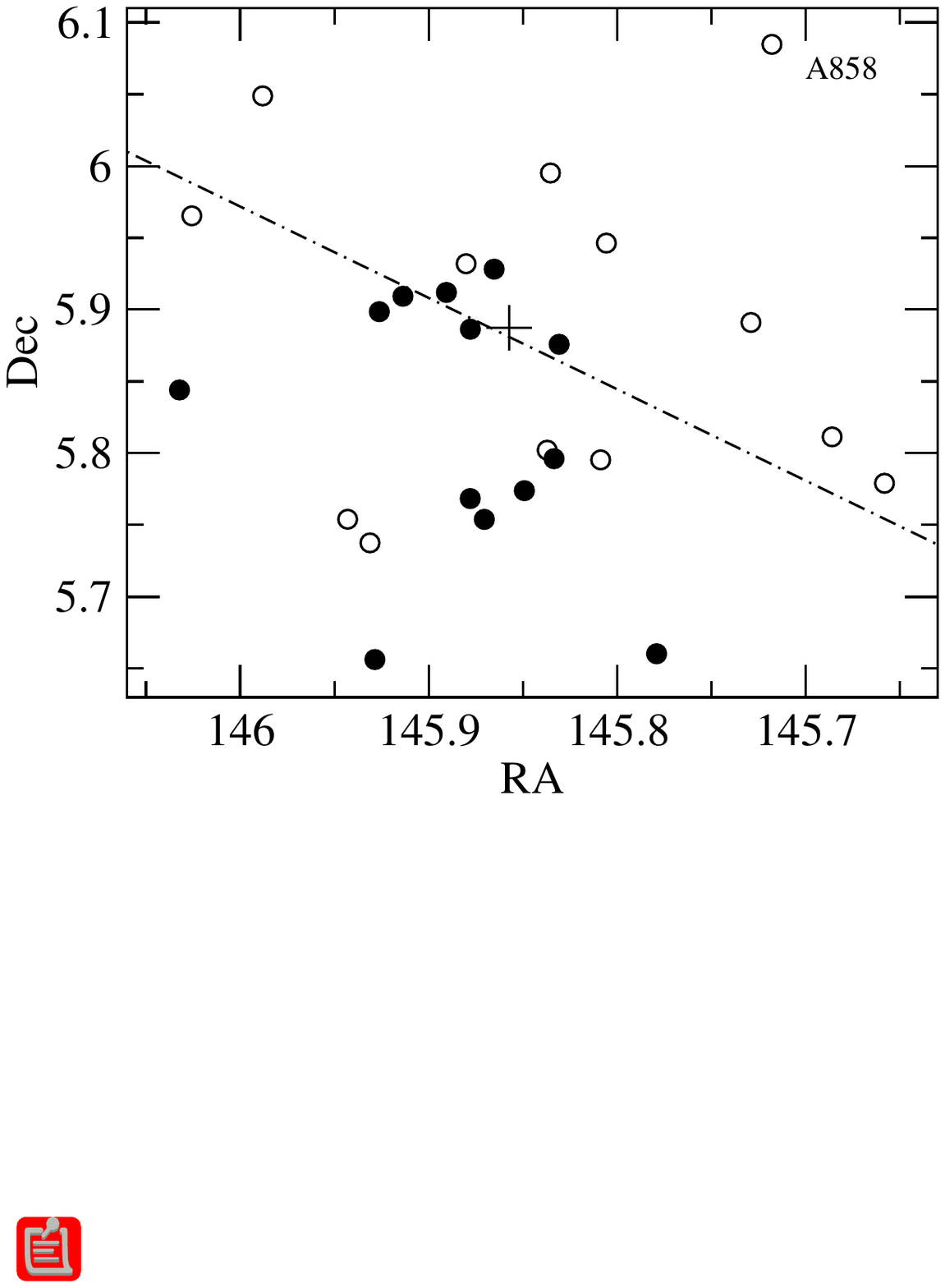}
\caption{The map of central region ($0.75R_A$) of the cluster A858.}
\end{figure}

\begin{figure}
\includegraphics[width=16cm]{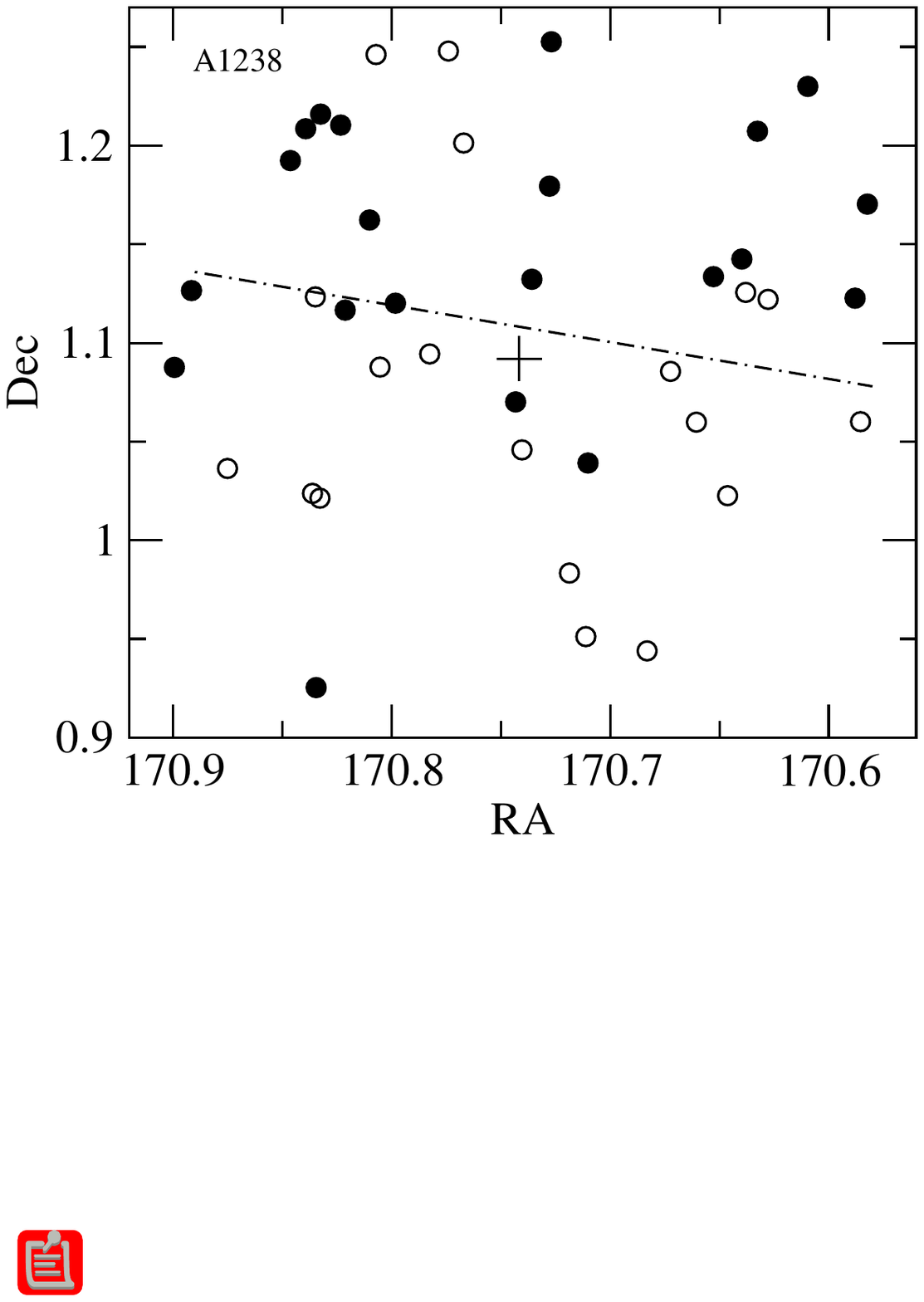}
\caption{The map of the central region ($0.5R_A$) of the rotating cluster A1238.}
\end{figure}

\begin{figure}
\includegraphics[width=16cm]{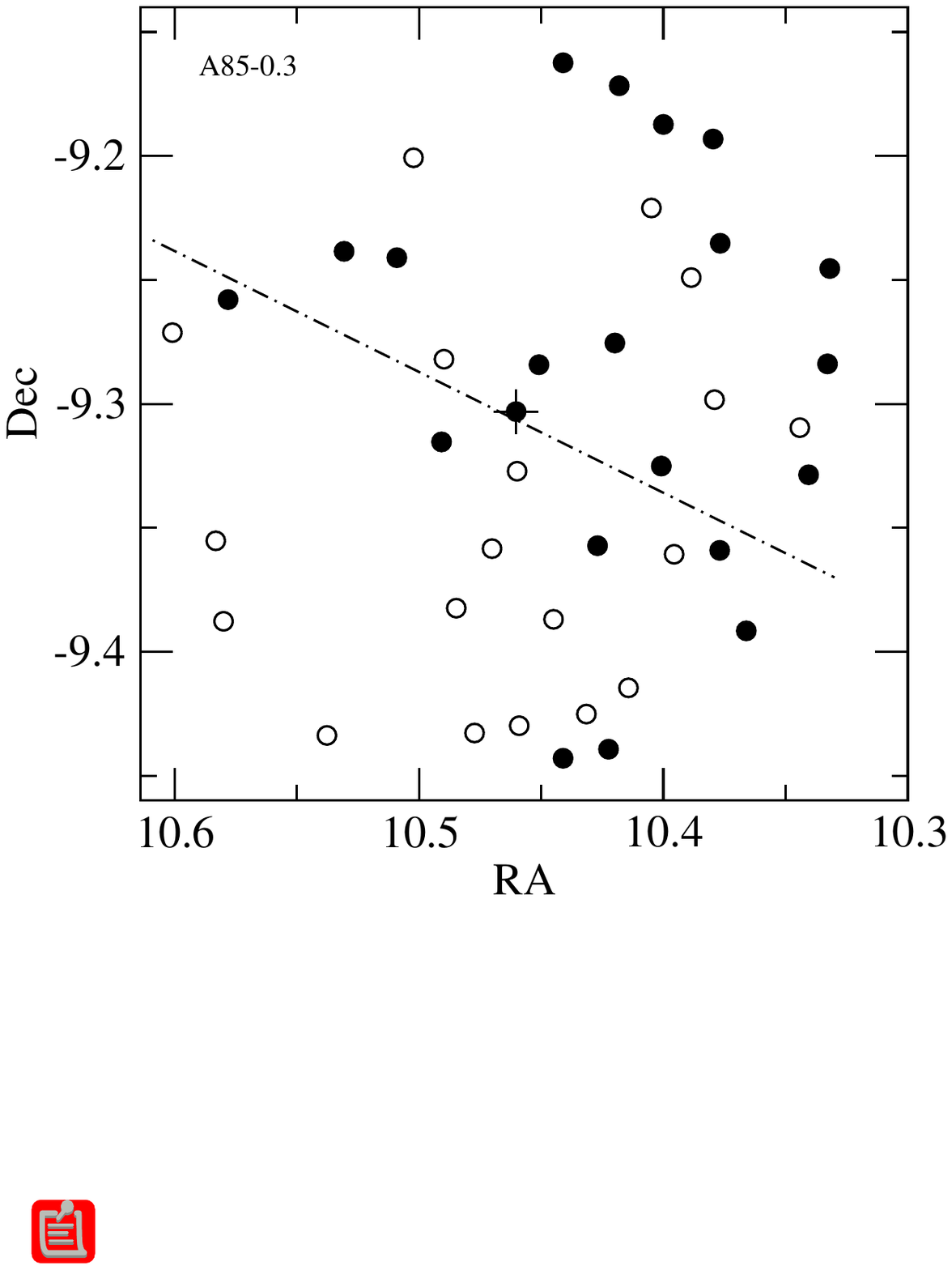}
\caption{The map of the central region ($0.30R_A$) of the rotating cluster A85.}
\end{figure}

\begin{figure}
\includegraphics[width=16cm]{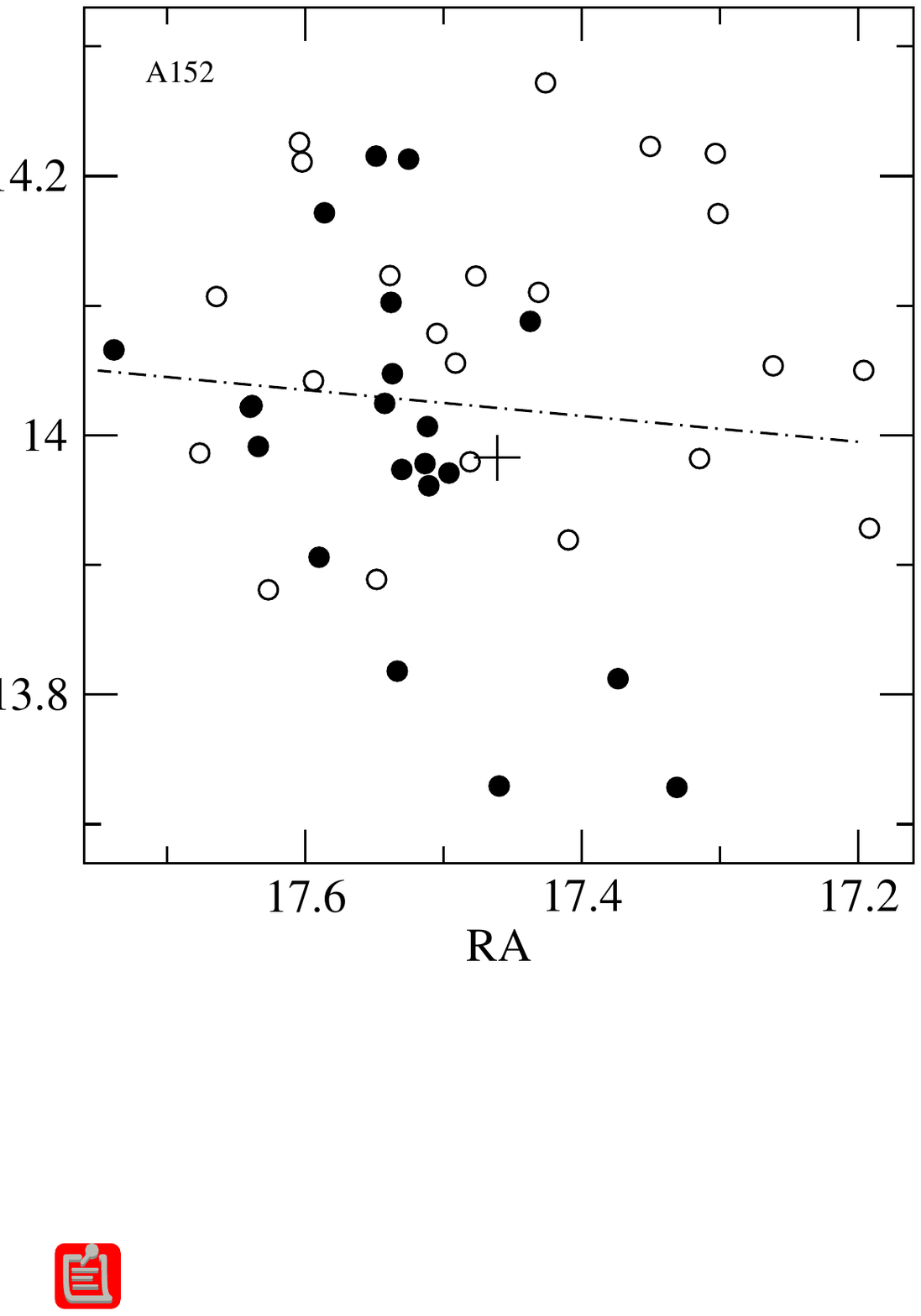}
\caption{The map of the central region ($0.60R_A$) of the rotating cluster A152.}
\end{figure}

\begin{figure}
\includegraphics[width=16cm]{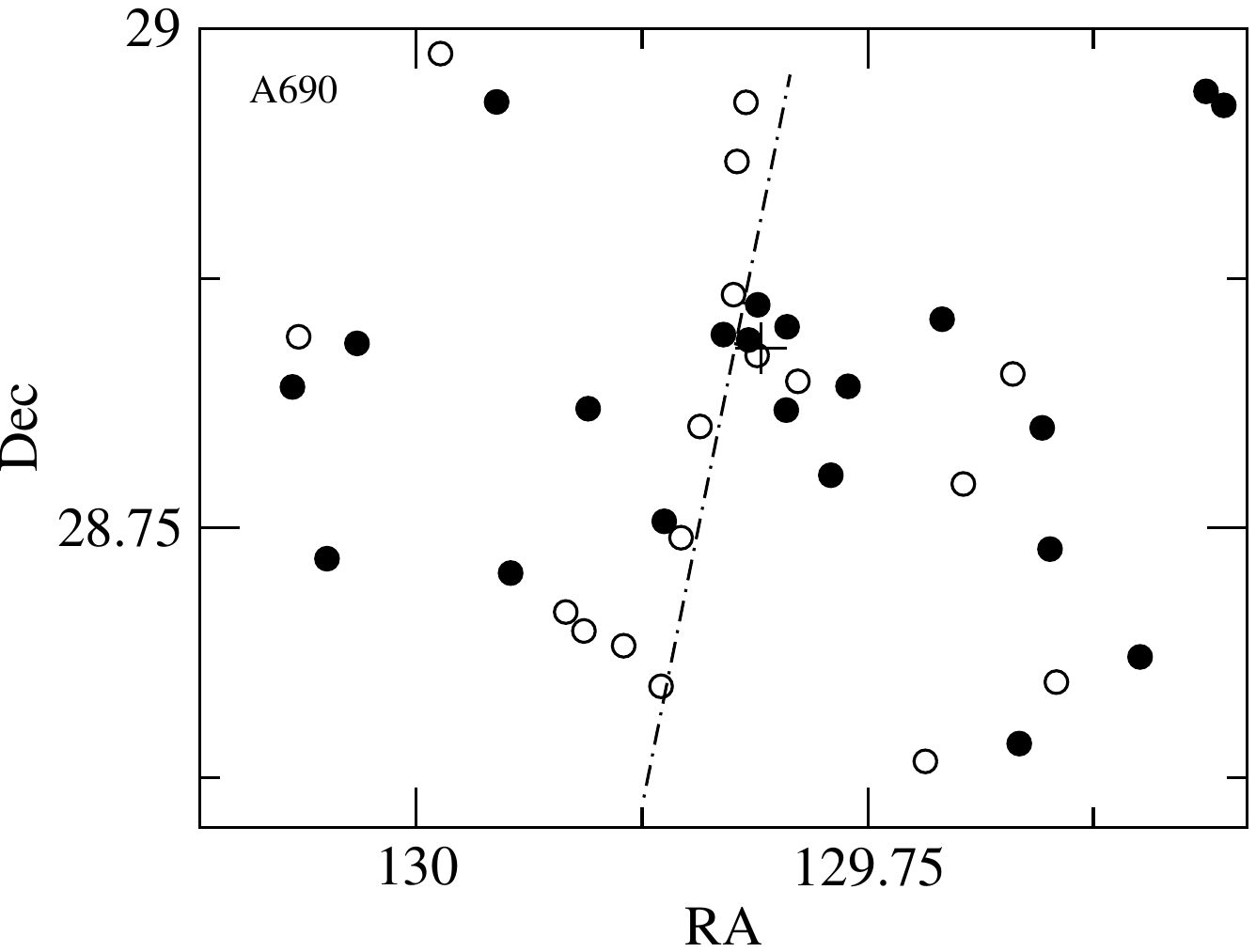}
\caption{The map of the central region ($0.75R_A$) of the rotating cluster A690.}
\end{figure}

\begin{figure}
\includegraphics[width=16cm]{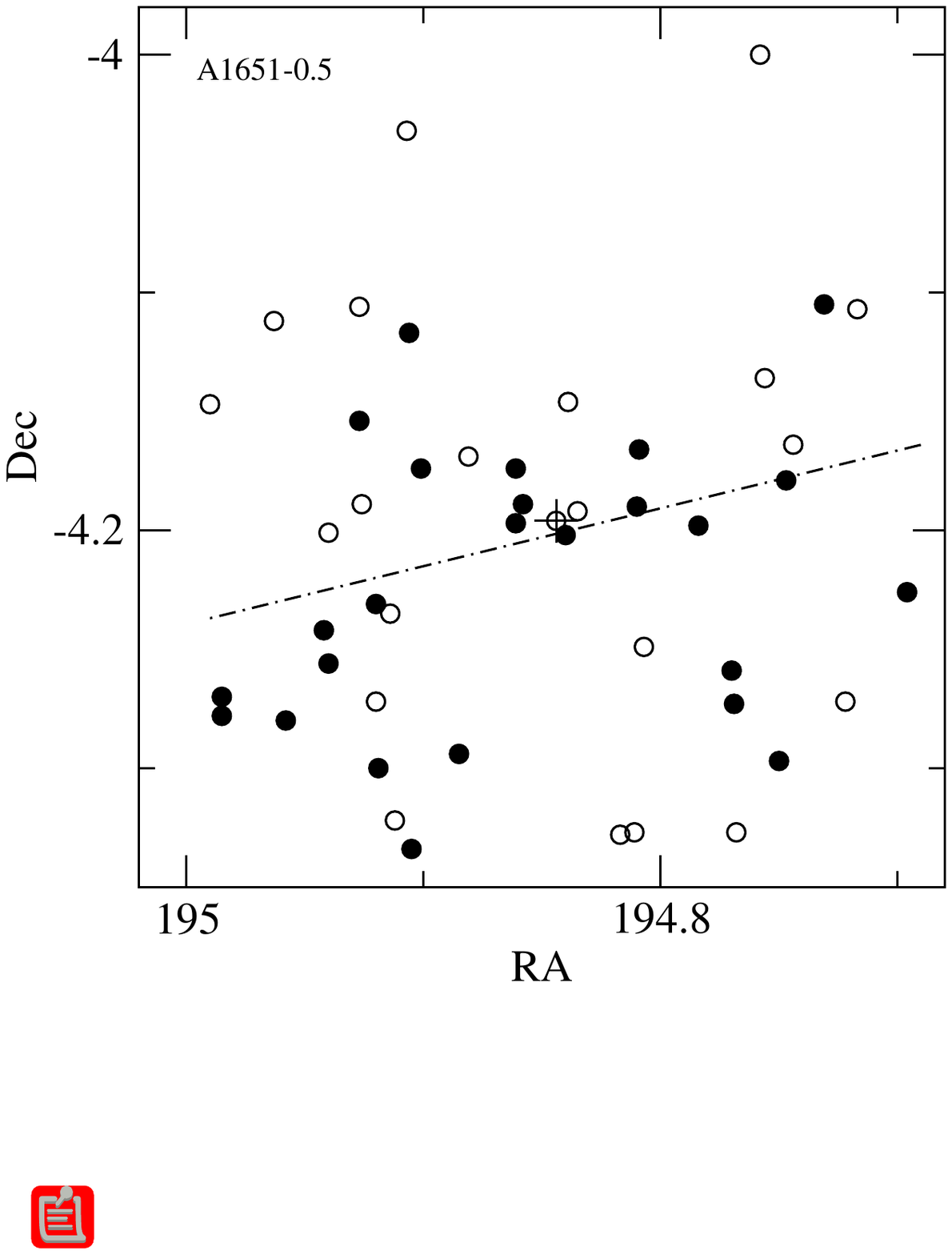}
\caption{The map of the central region ($0.5R_A$) of the rotating cluster A1651.}
\end{figure}

\begin{figure}
\includegraphics[width=16cm]{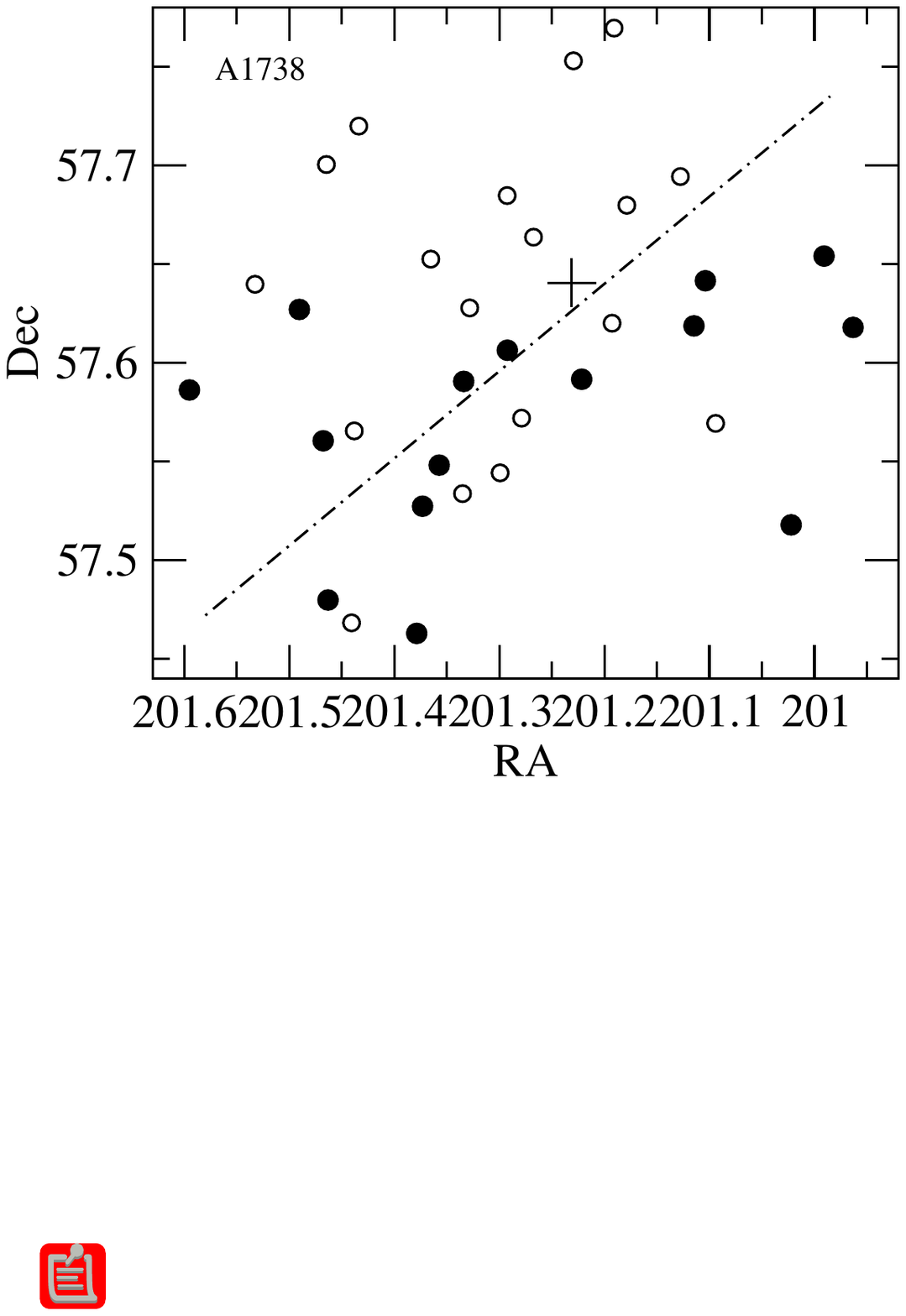}
\caption{The map of the central region ($0.75R_A$) of the rotating cluster A1738.}
\end{figure}

\begin{figure}
\includegraphics[width=16cm]{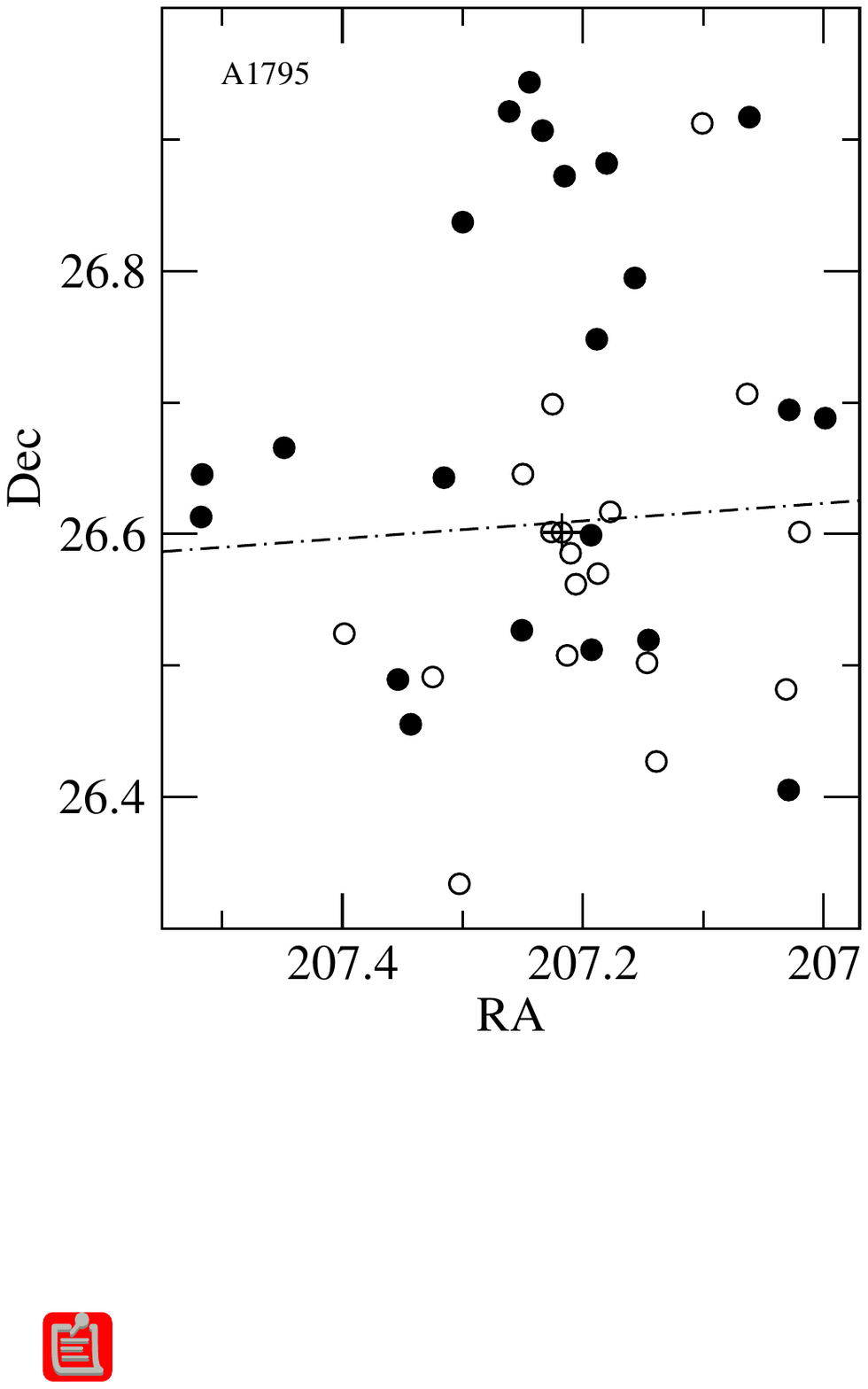}
\caption{The map of the central region ($0.40R_A$) of the rotating cluster A1795.}
\end{figure}

\begin{figure}
\includegraphics[width=16cm]{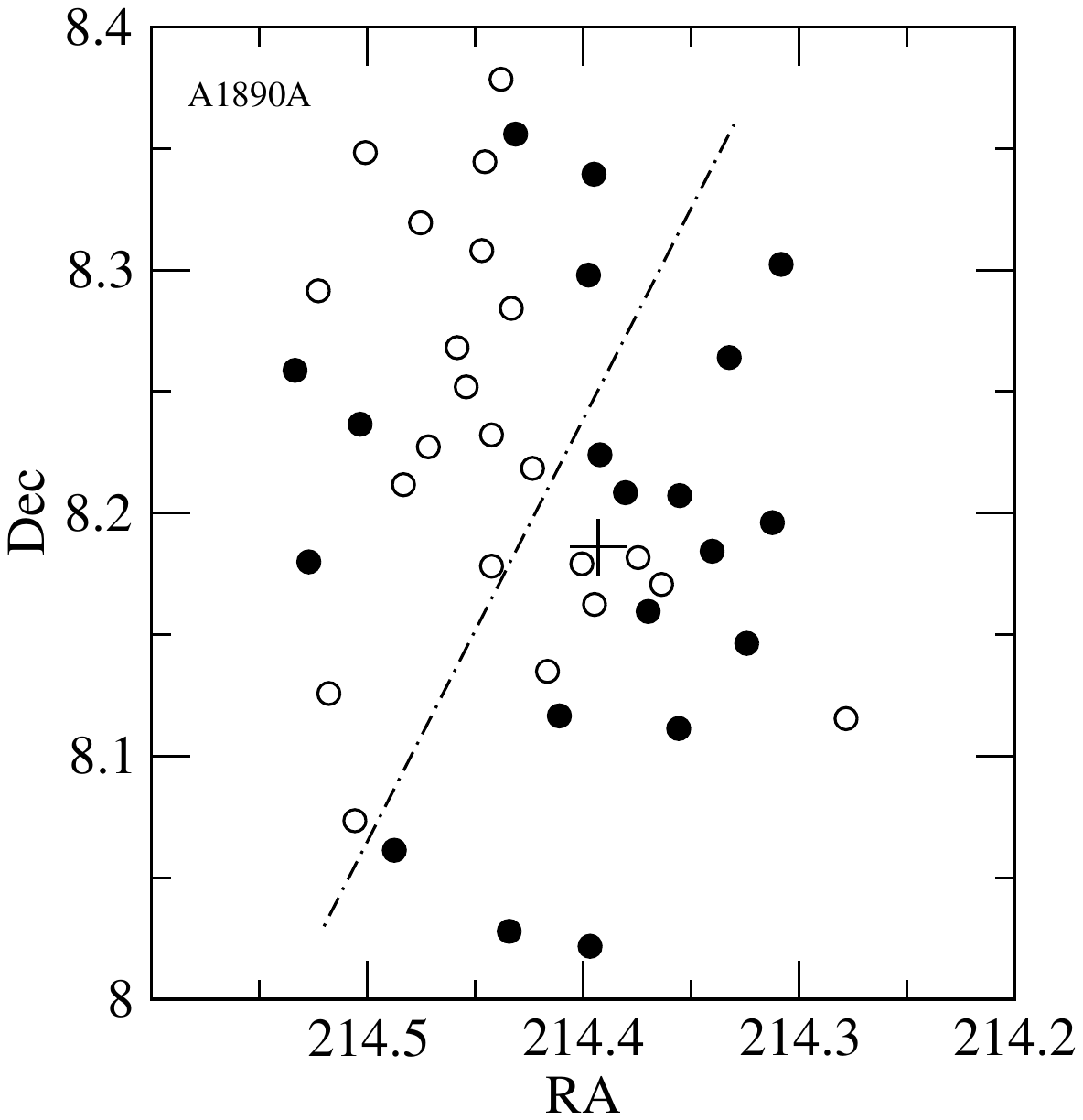}
\caption{The map of the central region ($0.5R_A$) of the rotating cluster A1890.}
\end{figure}

\end{document}